\documentclass[preprint]{aastex6}
\def\degpoint{\ifmmode ^{\rm{o}}\!. \else $^{\rm{o}}\!.$\fi}
\newcommand{\degrees}{$^{\rm{o}}$}
\newcommand{\ms}{\mbox{m\,s$^{-1}$}}

\newcommand{\Msun}{\mbox{$M_{\odot}$}}
\newcommand{\Rsun}{\mbox{$R_{\odot}$}}
\newcommand{\Mjup}{\mbox{$M_{\rm Jup}$}}

\newcommand{\Lsun}{\mbox{$L_{\odot}$}}

\newcommand{\gtsimeq}{\raisebox{-0.6ex}{$\,\stackrel
         {\raisebox{-.2ex}{$\textstyle >$}}{\sim}\,$}}

\usepackage{graphicx}
\usepackage{subfigure}

\begin{document}

\title{The Pan-Pacific Planet Search VII: The most eccentric planet 
orbiting a giant star}

\author{Robert A.~Wittenmyer\altaffilmark{1,2}, M.I. 
Jones\altaffilmark{3,4}, Jonathan Horner\altaffilmark{1,2}, Stephen R. 
Kane\altaffilmark{5}, J.P. Marshall\altaffilmark{1,6}, A.J. 
Mustill\altaffilmark{7}, J.S. Jenkins\altaffilmark{8}, P.A. Pena 
Rojas\altaffilmark{8}, Jinglin Zhao\altaffilmark{2}, Eva 
Villaver\altaffilmark{9}, R.P. Butler\altaffilmark{10}, Jake 
Clark\altaffilmark{1} }

\altaffiltext{1}{University of Southern Queensland, Computational 
Engineering and Science Research Centre, Toowoomba, Queensland 4350, 
Australia}
\altaffiltext{2}{Australian Centre for Astrobiology, UNSW Australia, 
Sydney 2052, Australia}
\altaffiltext{3}{European Southern Observatory, Alonso de C\'ordova 
3107, Casilla 19001, Santiago, Chile}
\altaffiltext{4}{ESO Fellow.}
\altaffiltext{5}{Department of Earth Sciences, University of California, 
Riverside, CA 92521, USA}
\altaffiltext{6}{Academia Sinica, Institute of Astronomy and 
Astrophysics, Taipei 10617, Taiwan}
\altaffiltext{7}{Lund Observatory, Department of Astronomy \& Theoretical 
Physics, Lund University, Box 43, SE-221 00 Lund, Sweden}
\altaffiltext{8}{Departamento de Astronom\'ia, Universidad de Chile, 
Casilla 36-D, Santiago, Chile}
\altaffiltext{9}{Departmento F\'isica Te\'orica, Facultad de Ciencias, 
Universidad Aut\'onoma de Madrid, Cantoblanco, 28049, Madrid, Espa\~na}
\altaffiltext{10}{Department of Terrestrial Magnetism, Carnegie 
Institution of Washington, 5241 Broad Branch Road, NW, Washington, DC 
20015-1305, USA}
\email{rob.w@usq.edu.au}

\shorttitle{PPPS VII: Most eccentric evolved-star planet}
\shortauthors{Wittenmyer et al.}


\begin{abstract}

\noindent Radial velocity observations from three instruments reveal the 
presence of a 4\,\Mjup\ planet candidate orbiting the K giant HD\,76920.  
HD\,76920b has an orbital eccentricity of 0.856$\pm$0.009, making it the 
most eccentric planet known to orbit an evolved star.  There is no 
indication that HD\,76920 has an unseen binary companion, suggesting a 
scattering event rather than Kozai oscillations as a probable culprit 
for the observed eccentricity.  The candidate planet currently 
approaches to about four stellar radii from its host star, and is 
predicted to be engulfed on a $\sim$100 Myr timescale due to the 
combined effects of stellar evolution and tidal interactions.

\end{abstract}

\keywords{planetary systems, stars: giants, stars: individual 
(HD\,76920), techniques: radial velocity }


\section{Introduction}

Prior to the dawn of the exoplanet era, astronomers felt that they 
understood how other planetary systems would look, and how the Solar 
system formed \citep[e.g.][and references therein]{Lissauer}. They 
expected to find rocky, telluric worlds, in the inner reaches of 
planetary systems, with gas giants further out. They expected planetary 
systems to be co-planar, or close to it, and that the planets they would 
find would move on low-eccentricity orbits. All of these ideas were 
based on our knowledge of the one planetary system that we knew at that 
time - the Solar system.

With the discovery of the first planet around a Sunlike star, 51 Peg 
\citep{51peg}, these assumptions began to fall around us. The first of 
many hot-Jupiters \citep[e.g.][]{HJ1,HJ3,HJ2}, 51 Pegasi b was the 
antithesis of our own planetary system - a giant, Jupiter-mass planet 
practically skimming the surface of its host star. And as the exoplanet 
era proceeded, new discoveries continued to shatter our old assumptions, 
revealing that the diversity of planetary systems is far greater than we 
could ever have imagined. Some systems contain planets whose orbits are 
highly inclined, or even retrograde, with respect to their host's 
equatorial plane \citep[e.g.][]{RM1,RM2,RM3}. Others contain planets far 
denser than those in the Solar system 
\citep[e.g.][]{Dense1,Dense2,Dense3}, or far fluffier 
\citep[e.g.][]{fluffy1,fluffy2,fluffy3}. Many systems contain planets 
unlike anything found around the Sun, with so-called `super-Earths' and 
`sub-Neptunes' proving to be common in the cosmos \citep{SN1,SN2,SN3, 
SN4}. There is even growing evidence of a large population of 
`free-floating planets', interstellar vagabonds roaming the depths of 
space \citep{FF1,FF2,FF3}. And then there are the eccentric planets - 
bodies moving on orbits more akin to the Solar system's comets than its 
planets \citep[e.g.][]{ecc2,ecc1,ecc3}

A significant fraction of these discoveries have relied on radial 
velocity observations, which are required in order to estimate the 
minimumn mass of newly discovered planets. In the first decade of the 
exoplanet era, the radial velocity technique was by far the most 
effective tool for the discovery of exoplanets, as well as their 
characterisation, and it remains the principal method by which systems 
that truly resemble the Solar system (with distant, massive planets) can 
be discovered, and thence the true frequency of Solar system analogues 
determined \citep[e.g.][]{etaearth,anal1,anal2,anal3}. Whilst the radial 
velocity technique is an excellent tool for the detection and 
characterisation of planets around Solar-type and late-type stars, it 
cannot be used to search for planets around massive, early-type stars. 
As such, whilst the occurance of planets around low mass stars is now 
becoming well-established out to relatively large orbital radii, the 
frequency and distribution of planets around more massive stars remains 
an open and fascinating question \citep[e.g.][]{retired1}.

In order to learn more about the occurrence and properties of planets 
around more massive stars, several teams have begun surveys of `retired 
A-stars'. Over the past few years, such surveys have begun to bear 
fruit, with a number of massive planets being found 
\citep[e.g.][]{retired2,retired3,jones11,n15,reffert15,33844}. As a 
result, we are now beginning to understand the relationship between 
stellar mass and the abundance of giant planets - with strong 
indications that giant planets are more efficiently formed around more 
massive stars \citep[e.g.][]{mal13, reffert15,retired5, ppps6}.

In this paper, we report the discovery of a highly eccentric planet 
orbiting the evolved star HD\,76920. In section 2, we detail our 
observations, and describe the stellar properties of HD\,76920. In 
section 3, we detail the orbit fitting process, and provide the 
parameters of the newly detected companion, before presenting and 
discussing our conclusions in section 4.


\section{Observations and Data Reduction}
\label{obs}

Observations of HD\,76920 were obtained with three different high 
resolution spectrographs, namely UCLES \citep{diego:90}, at the 3.9m 
Anglo-Australian Telescope, CHIRON \citep{toko13} installed at the 1.5m 
telescope in Cerro Tololo and FEROS \citep{kaufer99}, at the 2.2.m 
telescope in La Silla.  Both UCLES and CHIRON use an iodine cell, which 
is placed in the stellar light path, superimposing a rich absorption 
line spectrum, which is used to compute a precise wavelength reference 
\citep{val:95,BuMaWi96}.  On the other hand, FEROS is equipped with two 
fibres, to simultaneously record the stellar spectrum (in the science 
fibre) and a ThAr lamp spectrum (in the sky fibre), from which the 
nightly instrumental drift can be substracted \citep{baranne96}.  The 
UCLES data reduction and radial velocity computation method is described 
in \citet{BuMaWi96} and \citet{tinney01}.  The CHIRON data were 
extracted and calibrated with the pipeline offered by the CHIRON team, 
while the radial velocities were computed using the method described in 
\citet{jones17}.  Finally, the extraction and calibration of the FEROS 
data was performed with the CERES code \citep{brahm17}, and the radial 
velocities were obtained using the method presented in \citet{jones17}.

Stellar properties for HD\,76920 were derived from iodine-free template 
UCLES spectra with $R\sim\,60,000$, as described fully in \citet{ppps5}.  
In brief, spectroscopic stellar parameters were determined via a 
standard 1D, local thermodynamic equilibrium (LTE) abundance analysis 
using the 2013 version of MOOG \citep{sne73} with the ODFNEW grid of 
Kurucz ATLAS9 model atmospheres \citep{cas03}.  Complete stellar 
parameters from \citet{ppps5} and other literature sources are given in 
Table~\ref{stellarparams}.


\section{Data Analysis and Companion Parameters}

\subsection{Orbit fitting}

The AAT/UCLES data for HD\,76920 are relatively constant, save for two 
observations nearly 300\ms\ higher on 2010 Jan 29/30.  These spectra 
were of similar S/N to the others for this target, and the observing 
conditions were typical, ruling out the possibility of observer error or 
systematic errors (e.g. scattered moonlight) for the aberrant 
velocities.  Reprocessing the spectra through an independent reduction 
and Doppler pipeline gave similar results, ruling out errors in data 
reduction or barycentric correction.  We attempted a single 
highly-eccentric planet fit using the genetic algorithm employed by our 
team for other Pan-Pacific Planet Search (PPPS) datasets with sparse 
sampling \citep[e.g.][]{47205, 121056, ppps6}.  We searched a period 
range of 100-1000 days and allowed eccentricities up to $e=0.9$, running 
for 50,000 iterations (about $10^7$ possible configurations).  The 
best-fit solution had a high eccentricity and a period of $\sim$420 
days.

\begin{figure}
\gridline{\fig{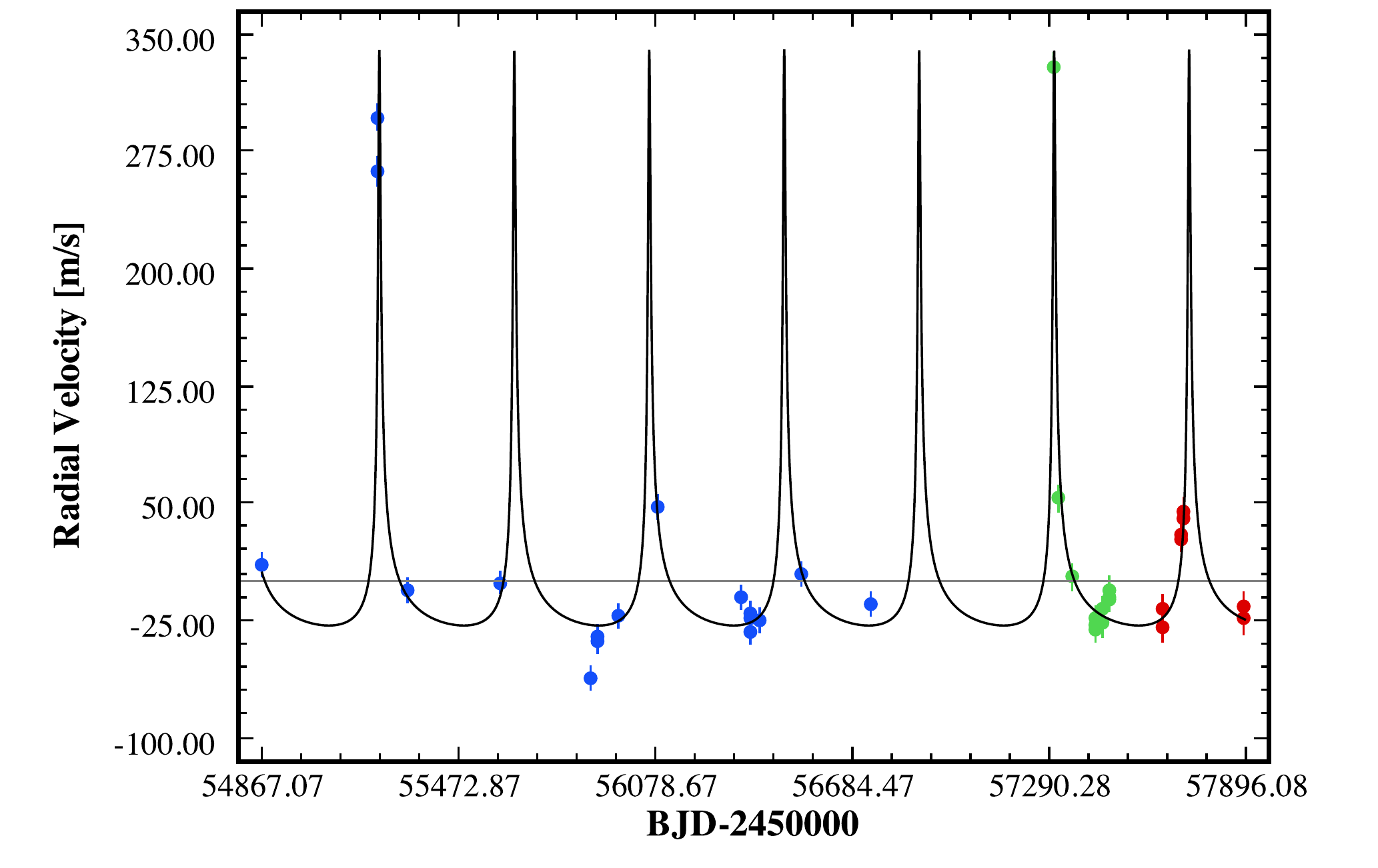}{0.5\textwidth}{(a)}
          \fig{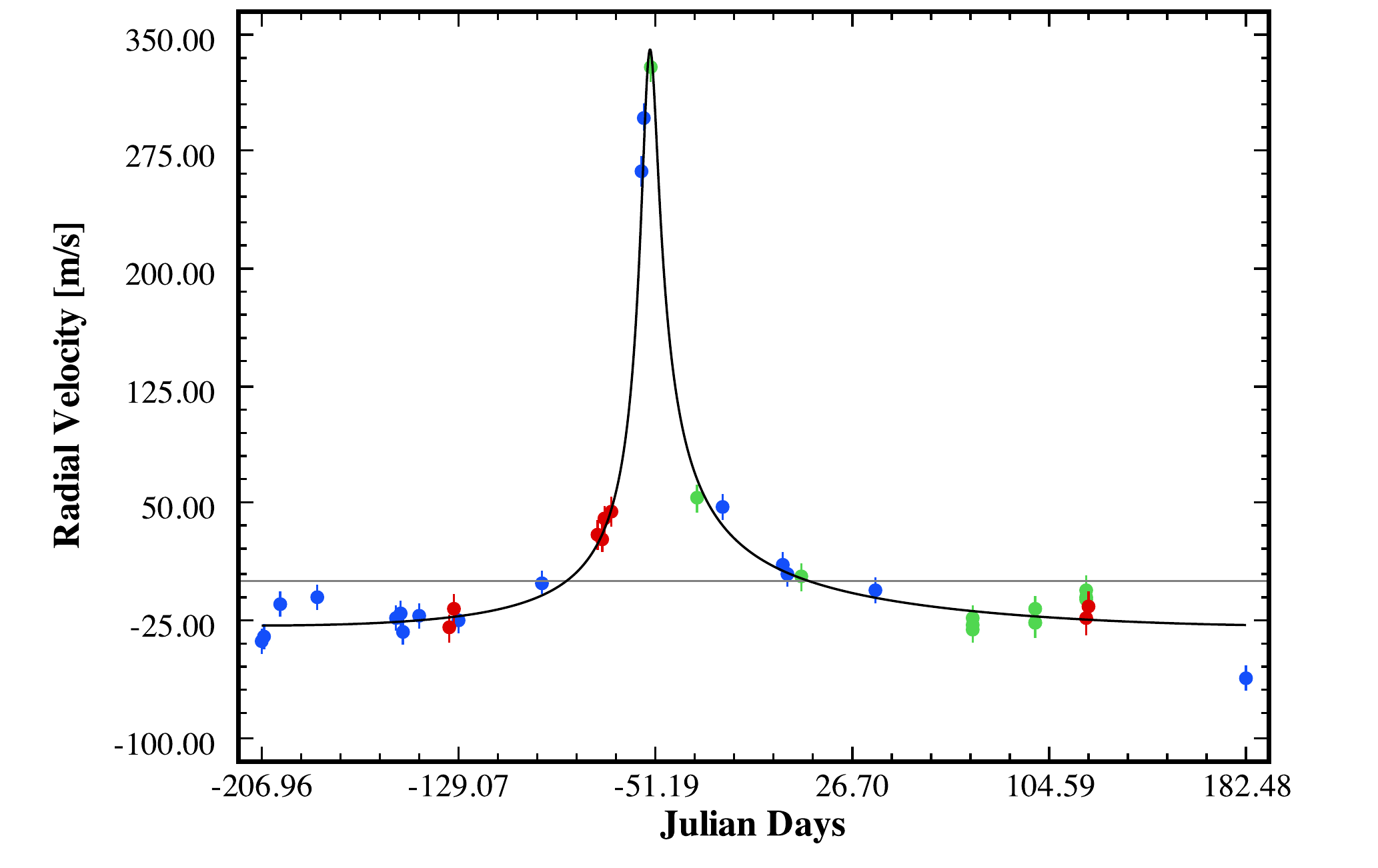}{0.5\textwidth}{(b)}}
\caption{Data and Keplerian fit for HD\,76920b (AAT -- blue, CHIRON -- 
green, FEROS -- red).  Error bars include 7\,\ms\ of jitter added in 
quadrature.  The rms about this fit is 9.74\,\ms.  Right: Same, but 
phase folded on the orbital period $P=415.4$ days. }
\label{fitplot}
\end{figure}

\begin{figure}
\gridline{\fig{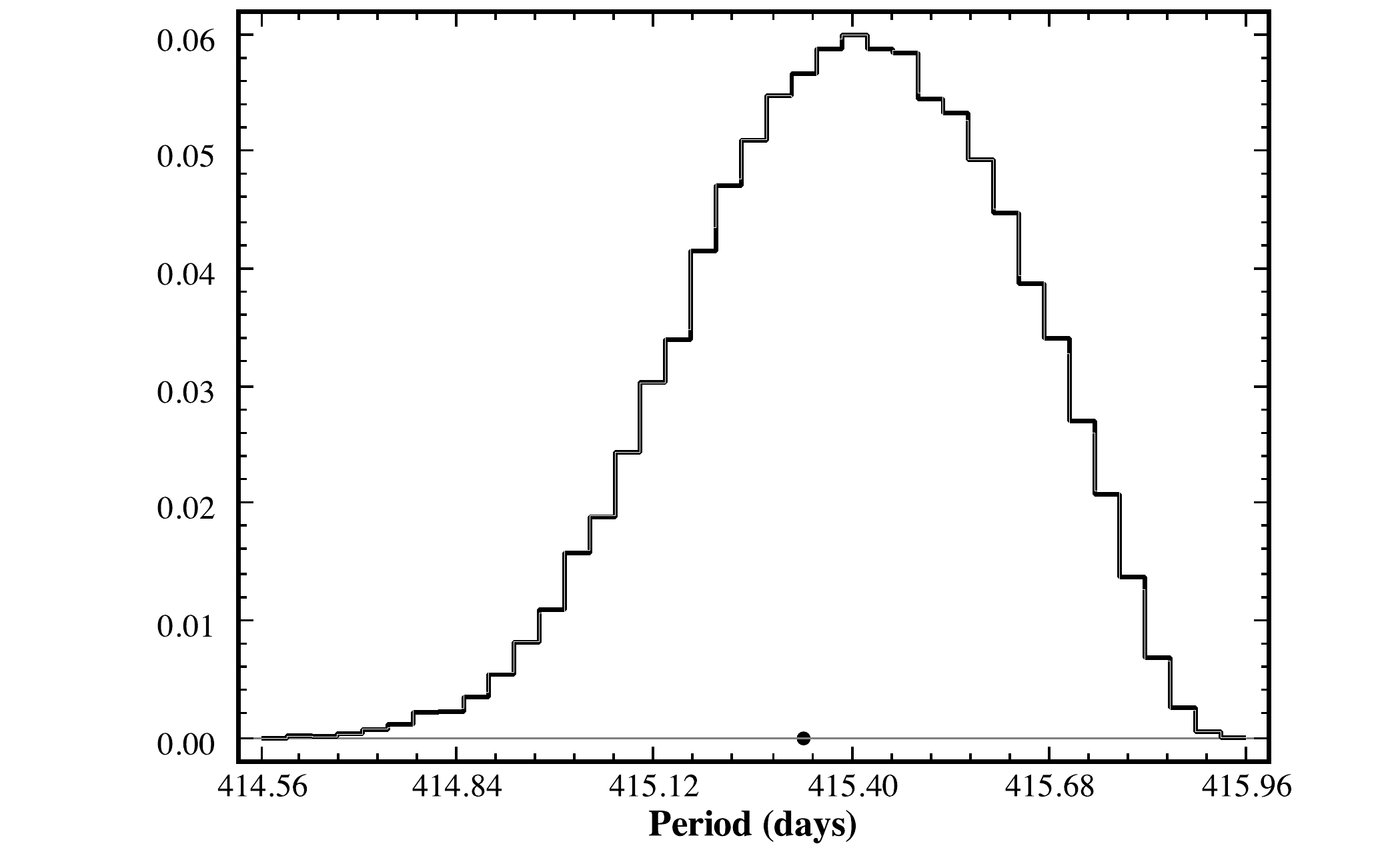}{0.5\textwidth}{(a)}
          \fig{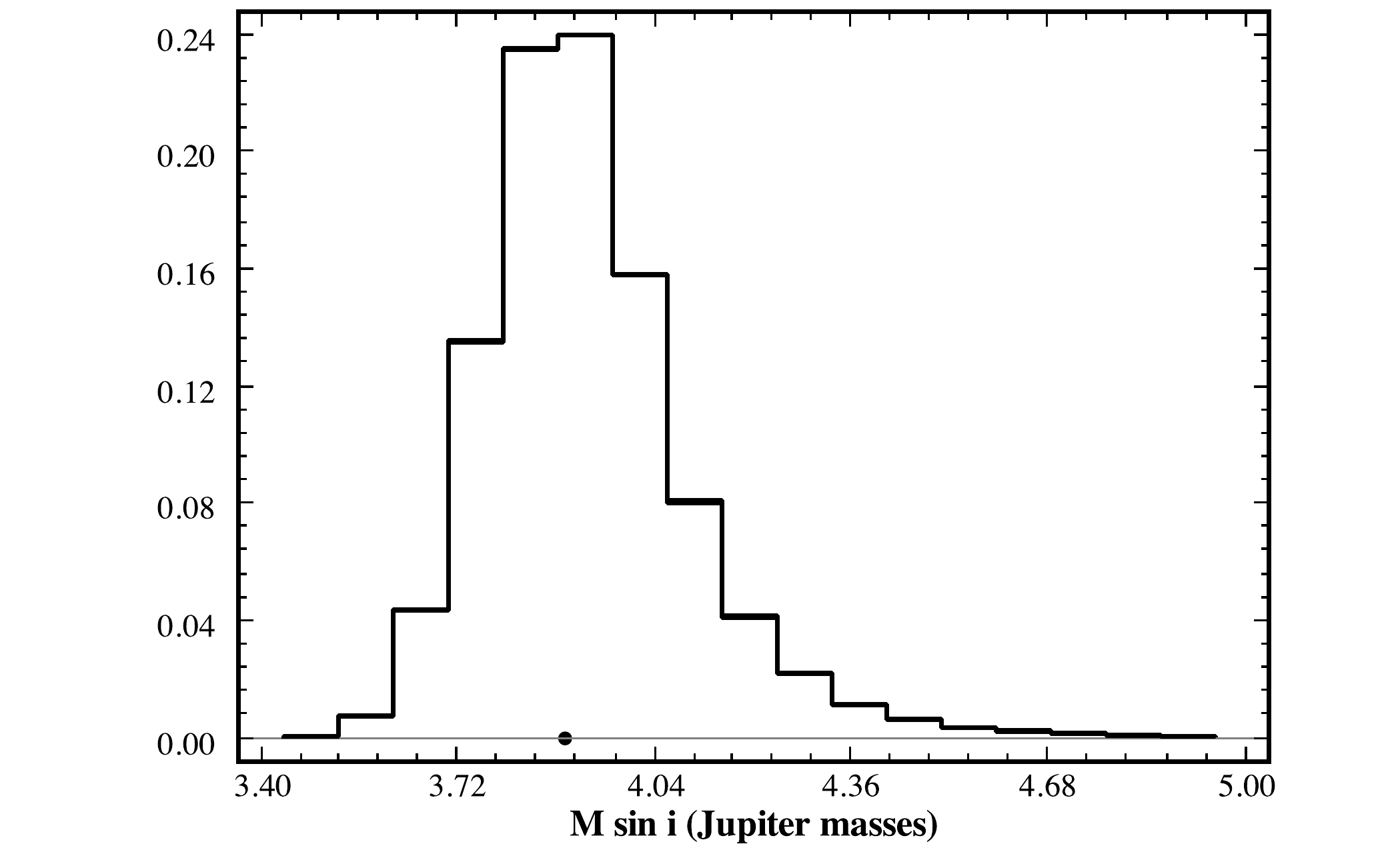}{0.5\textwidth}{(b)}}
\gridline{\fig{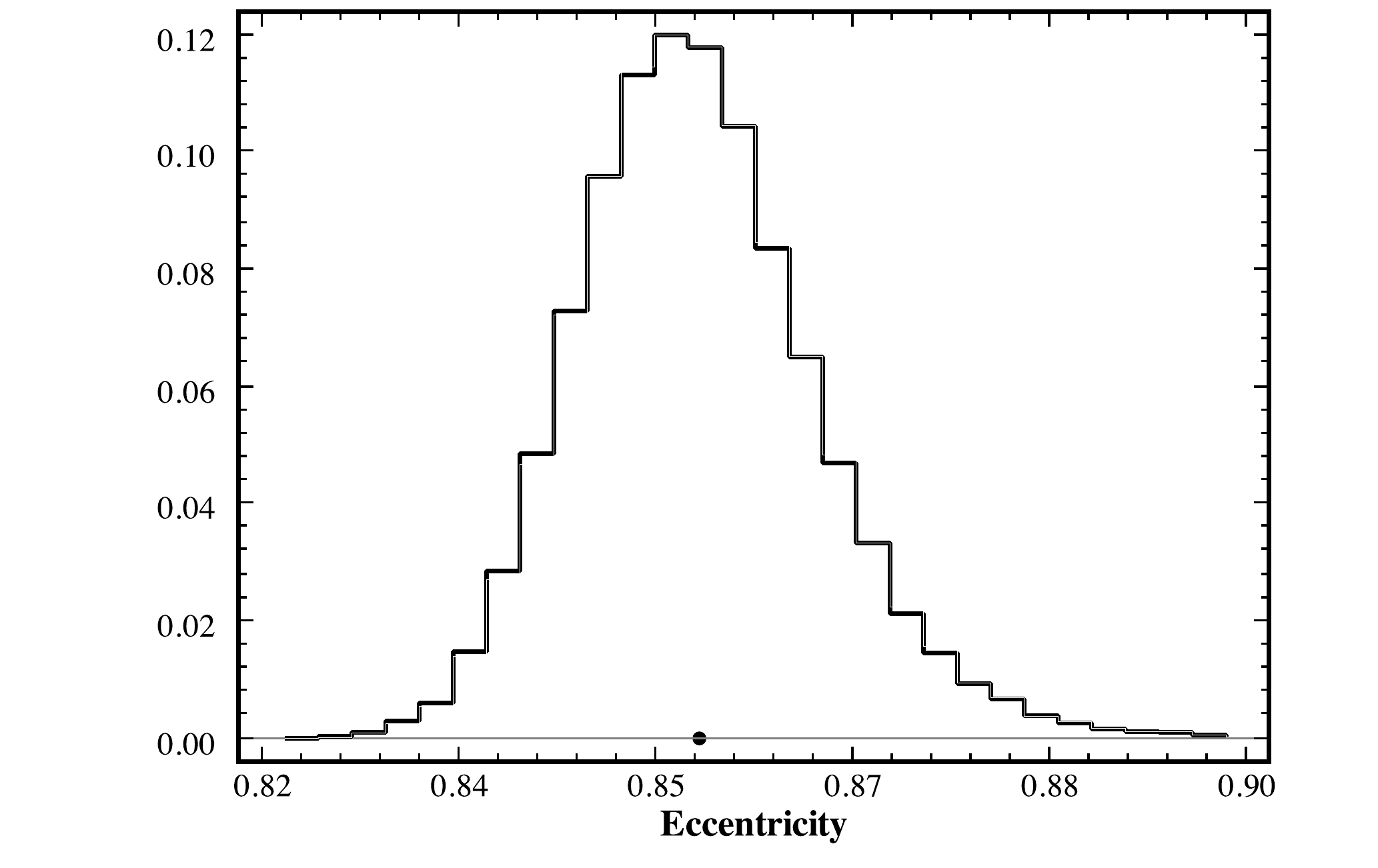}{0.5\textwidth}{(c)}
          \fig{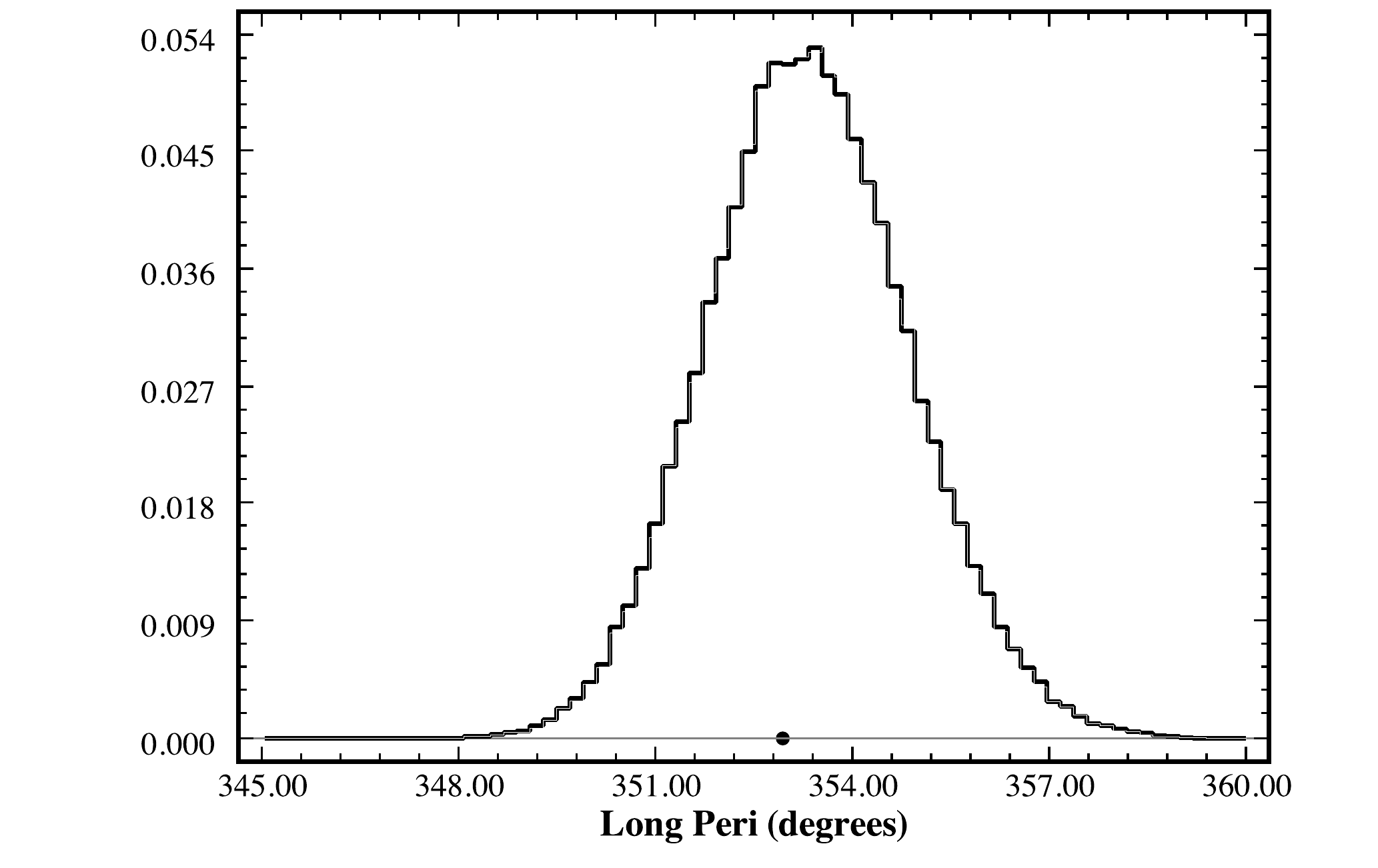}{0.5\textwidth}{(d)}}
\gridline{\fig{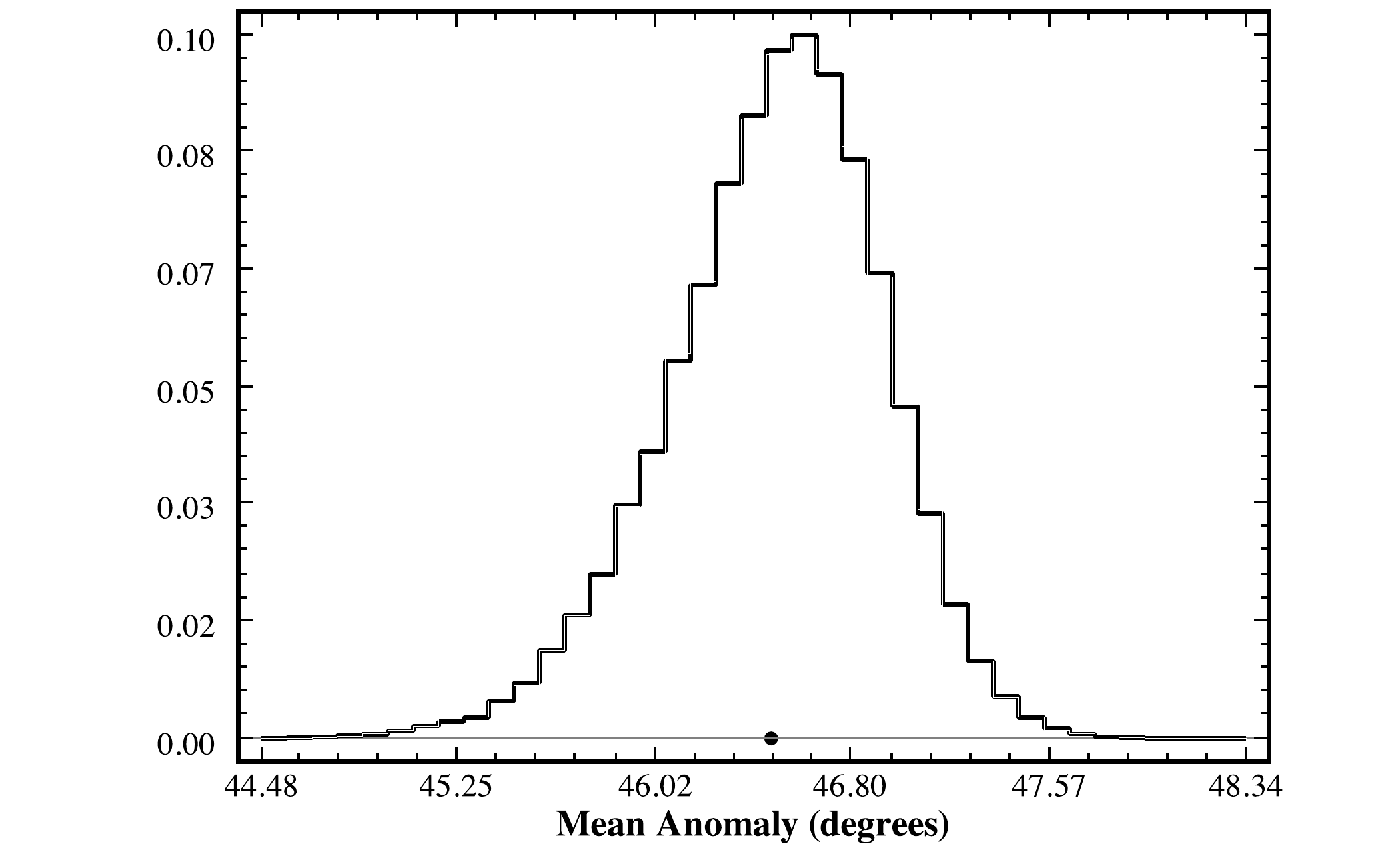}{0.5\textwidth}{(e)}}
\caption{Posterior distributions from MCMC analysis of the combined data 
for HD\,76920. }
\label{posteriors}
\end{figure}

Our team obtained access to CHIRON in 2015 for follow-up of interesting 
candidates from the PPPS, and HD\,76920 was put in the queue to catch 
the next predicted large velocity excursion of 2015 October.  We were 
fortunate to catch the peak and the drop-off, confirming the period at 
415 days.  FEROS observations were also added and corroborated the orbit 
fit obtained from AAT and CHIRON.  For the final fitting process, we 
used the Keplerian model in the \textit{Systemic Console} version 2.2000 
\citep{mes09}.  For all orbit fitting, 7\,\ms\ of jitter (excess 
white noise) has been added in quadrature to the internal instrumental 
uncertainties of each data set.  This estimate is derived from the 
velocity scatter of 37 stable stars in the PPPS as first described in 
\citet{33844}.  While correlated noise is known to affect radial 
velocity data \citep{baluev13, feng17}, we find that a white noise model 
is favoured for all three data sets.  The Bayes factors for a 
first-order moving average noise model (``MA(1)'') are as follows: AAT 
-- 2.3; FEROS -- 1.5; CHIRON -- -1.7.  The data and best fit model are 
shown in Figure~\ref{fitplot}.  Uncertainties in the system parameters 
were obtained with the Markov Chain Monte Carlo tool within 
\textit{Systemic}.  The normalised probability density functions for key 
parameters, resulting from a chain of $10^7$ steps, are shown in 
Figure~\ref{posteriors}.  The best fit to the data results in a planet 
with $P=415.4\pm$0.2 days, $m$~sin~$i=3.93\pm$0.15\,\Mjup, and 
$e=0.856\pm$0.009 (Table~\ref{planetparams}).  There is no evidence for 
additional Keplerian signals or velocity trends that might indicate a 
distant massive companion, as shown in Figure~\ref{resids}.

\subsection{Bayesian approach}

As a further test of the validity of the detected signal in the 
timeseries, we also ran the Exoplanet Mcmc Parallel tEmpering Radial 
velOcity fitteR (EMPEROR; Jenkins \& Pena 2017, in prep 
\footnote{https://github.com/ReddTea/astroEMPEROR}) code to determine if 
the parameter estimations were robust.  EMPEROR employs thermodynamic 
integration methods \citep{gregory05} following an affine invariant MCMC 
engine, performed using the \textsc{emcee} package \citep{fm13} in Python.  
Correlated noise in the measurements are taken care of within EMPEROR by 
using a first-order moving average model, an approach that has been 
shown to robustly detect small amplitude signals with various 
morphologies and across numerous radial velocity data sets 
\citep[e.g.][]{tuomi13, jenkins13, jenkins14, jenkins17}.  Model 
selection is performed automatically by EMPEROR, whereby a Bayes Factor 
of 5 is required, a threshold probability of 150 that the more complex 
model is favoured over the less complex one.  The code also 
automatically determines of the signal parameters like the period and 
amplitude are statistically significantly different from zero, and some 
basic priors are applied to those parameters whereby all are assumed to 
be flat except for the eccentricity and jitter priors that are folded 
gaussian and Jeffries priors, respectively. 

Under the aforementioned constraints, EMPEROR found two statistically 
significant signals in the radial velocity timeseries.  The primary 
signal was found to be the planet signal at 415.59$^{+0.19}_{-0.22}$ 
days, with an amplitude and eccentricity of 177.5$^{+6.4}_{-3.8}$\,\ms\ 
and 0.859$^{+0.005}_{-0.005}$, respectively. This result highlights the 
robustness of the planet detection result, showing that a long chain 
($\times 10^{7}$) MCMC analysis to probe the posterior parameter space, 
along with Bayesian selection criteria, can place hard constraints on the 
planet signal detection and the orbital characteristics of the planet.  
For completeness, we show a corner plot of the posterior 
distribution in Figure~\ref{corner}.


The EMPEROR analysis found a secondary signal which we discuss 
briefly here.  A comparison of evidence is given in Table~\ref{bics}.  
This second signal was found with a period of 28.4$^{+0.04}_{-0.57}$ days 
and an amplitude of 11.1$^{+1.5}_{-1.5}$\,\ms.  The amplitude is at the 
level expected for the jitter of HD\,76920: allowing the excess white 
noise (``jitter'') to vary in the single-planet model, we obtained the 
following for each instrument: AAT -- 10.4$\pm$1.5\,\ms, CHIRON -- 
8.3$\pm$1.6\,\ms, FEROS -- 2.9$\pm$1.9\,\ms.  Hence, this secondary 
signal is likely to originate from an intrinsic stellar process; 
attributing the periodicity of the secondary signal to a planet implies 
an orbit which crosses that of HD\,76920b.  Crossing orbits are almost 
certainly a recipe for dynamical disaster, with catastrophic 
instabilities occurring on timescales of a few years 
\citep[e.g.][]{HUAqr, QSVir, NSVS, hinse14}.

\subsection{Stellar activity}

As a matter of course for new planet discoveries, we searched for 
activity-related signals in the spectra and publicly available 
photometry.  Examination of 8.8 years (1403 epochs) of All-Sky Automated 
Survey (ASAS) photometry \citep{asas} shows no periodicities of 
significance near the planet's orbital period (Figure~\ref{photpgram}).  
The ASAS $V$ band photometry has a mean value of 7.827$\pm$0.013 mag.  
We also investigated the variability in the H$\alpha$ line for our 17 
AAT spectra.  Variable levels of chromospheric activity can produce 
changes in the level of line profile reversal in some line cores, 
resulting in changes to the line centroid and hence the measured radial 
velocity \citep{Martinez2010}.  These effects will also produce changes 
in the line's equivalent width (EW), and so measurement of the EW can 
provide an indicator of the presence of activity-induced radial velocity 
variations \citep{robertson14}.  Figure~\ref{activity} shows the stacked 
spectra of HD\,76920 and the radial velocity as a function of the 
H$\alpha$ EW.  No correlation is evident: the two velocity extrema, 
obtained on consecutive days, have quite different EW, one of which is 
consistent with the EW of the remaining spectra.  Lest this 
discrepancy raise concerns about the candidate planetary signal, we 
perform one final test: Figure~\ref{76920bAbides} shows the periodogram 
of our data with all three ``high'' velocities removed.  The highest 
peak remains at 413 days with a false-alarm probability of 0.6\%.



\section{Discussion and Conclusions}

\begin{figure}
\includegraphics[width=1.0\textwidth]{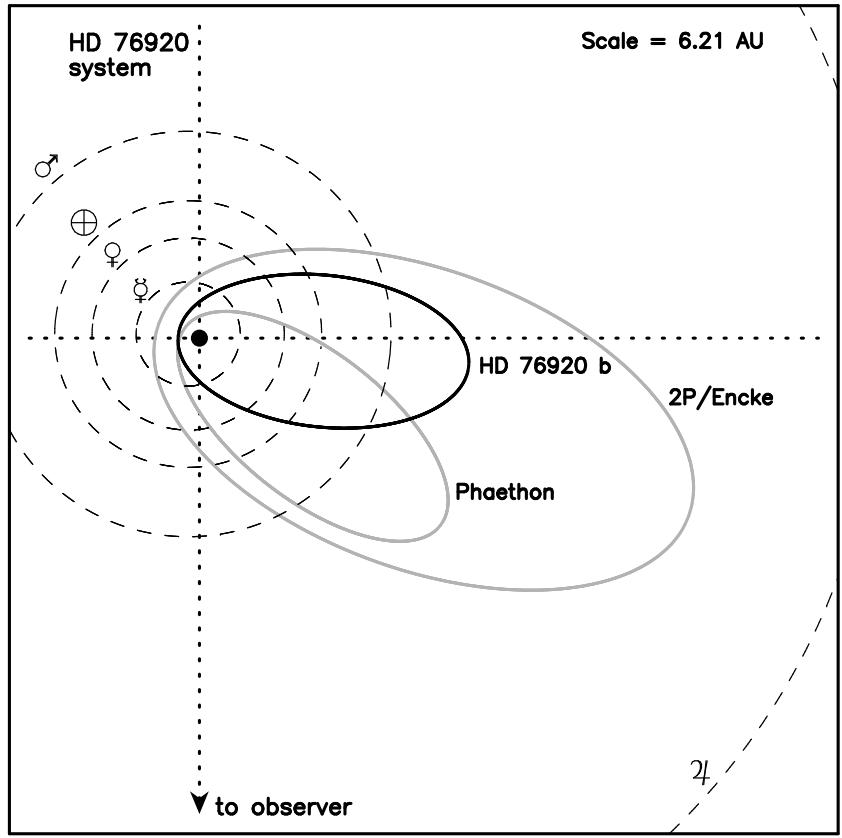}
\caption{Orbit of HD\,76920b, oriented properly and overlaid with the 
Solar System inner planets' orbits to scale.  Comet 2P/Encke and 
asteroid 3200 Phaethon are shown as examples of comparably eccentric 
Solar system bodies. }
\label{orbit}
\end{figure}

With $e=0.85$, HD\,76920b claims the title of the most eccentric planet 
known to orbit a giant star (i.e. with log $g<$3.5).  The previous 
record holder, iota Dra b, has $e=0.71$ \citep{butler06}.  To illustrate 
how extreme the orbit of HD\,76920b is, and how dramatically different 
it is to the planets in our own Solar system, it is useful to plot the 
planet's orbit alongside the inner Solar system. In Figure~\ref{orbit}, 
we show how the orbit of HD\,76920b compares to those of the telluric 
planets (Mercury, Venus, Earth and Mars), shown to scale. In addition, 
we include two of the Solar system's most famous small bodies - comet 
2P/Encke \citep[the parent of the Beta Taurid and Taurid meteor 
streams][]{Ash91} and asteroid 3200 Phaethon \citep[the parent of the 
Geminid meteor stream][]{Wil93}. Both these objects move on dynamically 
unstable orbits, and are only transient visitors to the inner Solar 
system. Comet Encke is a Jupiter-family comet, and was most likely 
injected to its current orbit from the Centaur population - icy bodies 
beyond the orbit of the giant planet \citep[e.g.][]{Hor3,Hor4,Lev6}. 
3200 Phaethon is a near-Earth asteroid, with an origin in the asteroid 
belt, interior to the orbit of Jupiter \citep[e.g.][]{Del10}. In both 
cases, it is clear that the objects did not form on their current 
orbits, but were instead transferred there from more distant, more 
circular orbits as a result of a lengthy series of gravitational 
perturbations. Where non-gravitational or secular perturbations are 
involved (as is the case for both comet 2P/Encke and asteroid 3200 
Phaethon), it is possible for the perturbed body to `decouple' from the 
more distant perturber, such that it no longer undergoes periodic close 
encounters that can dramatically alter its orbit. Given the tidal 
interactions that are likely occurring between HD\,76920b and its host 
star, there is clearly the potential for a similar process to be 
occurring in the HD\,76920 system - with the newly discovered planet 
having tidally decoupled from a distant perturber and then injected 
to its current highly eccentric orbit.

\begin{figure}
\includegraphics[width=1.0\textwidth]{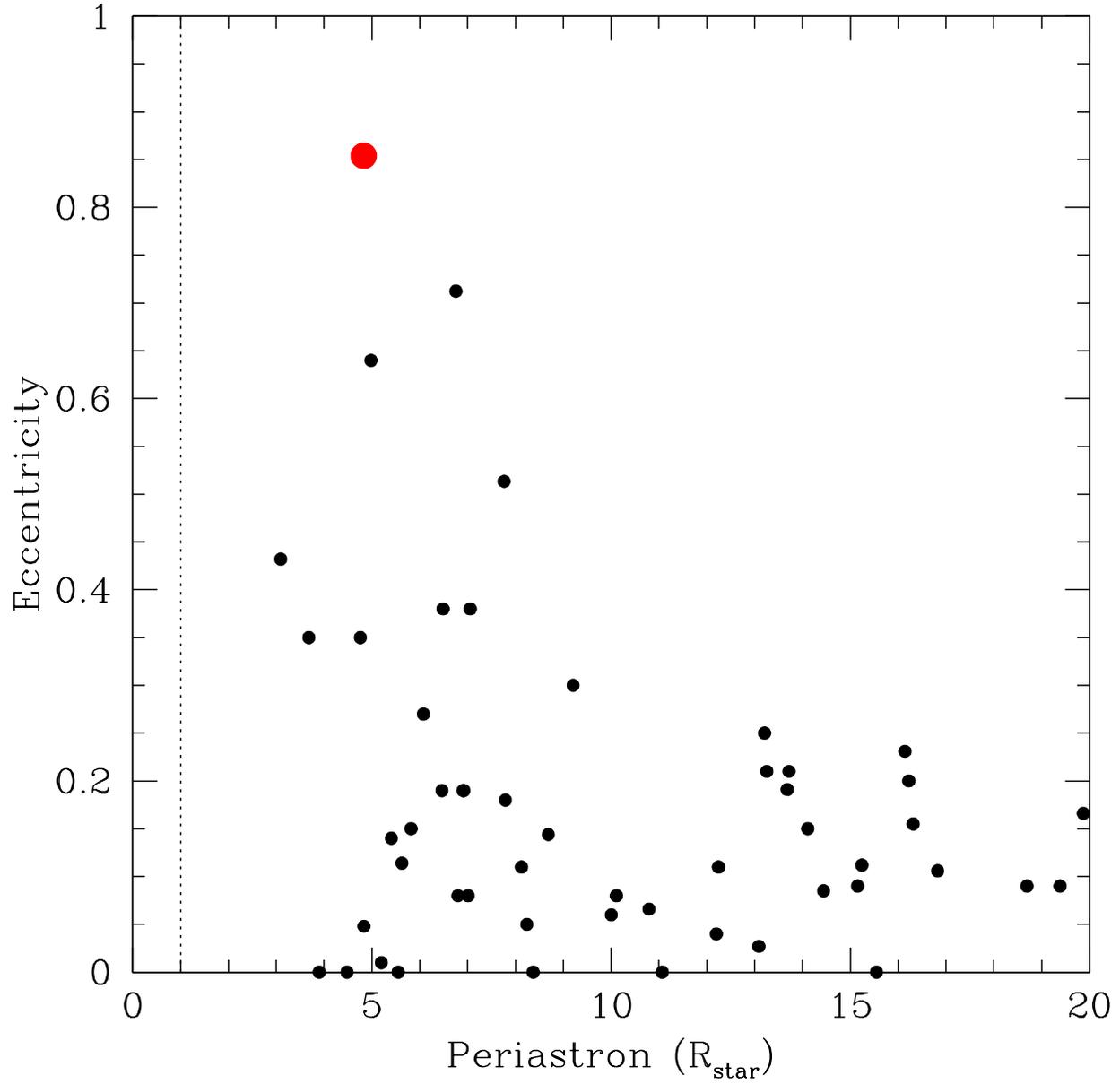}
\caption{Orbital eccentricity versus the planet's periastron distance, 
in terms of each planet's host-star radius, for 116 confirmed planets 
orbiting giant stars (log $g<$3.5). HD\,76920b, the most eccentric such 
planet, is shown as a large red point. }
\label{howclose}
\end{figure}

\begin{figure}
\includegraphics[width=1.0\textwidth]{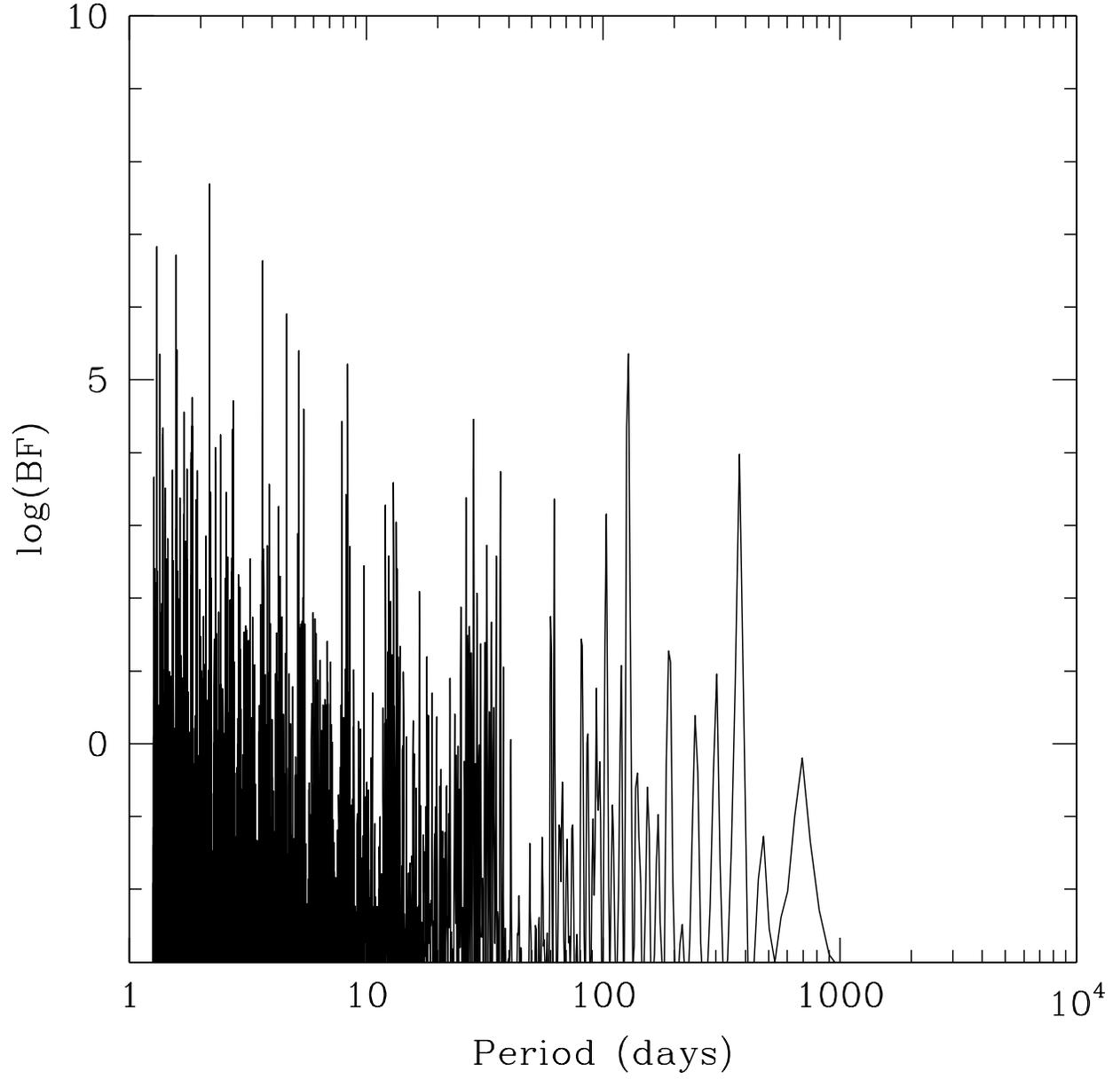}
\caption{Bayes Factor periodogram of the residuals to our 1-planet fit. 
No further periodicities of interest are evident. }
\label{resids}
\end{figure}

\begin{figure}
\includegraphics[width=1.0\textwidth]{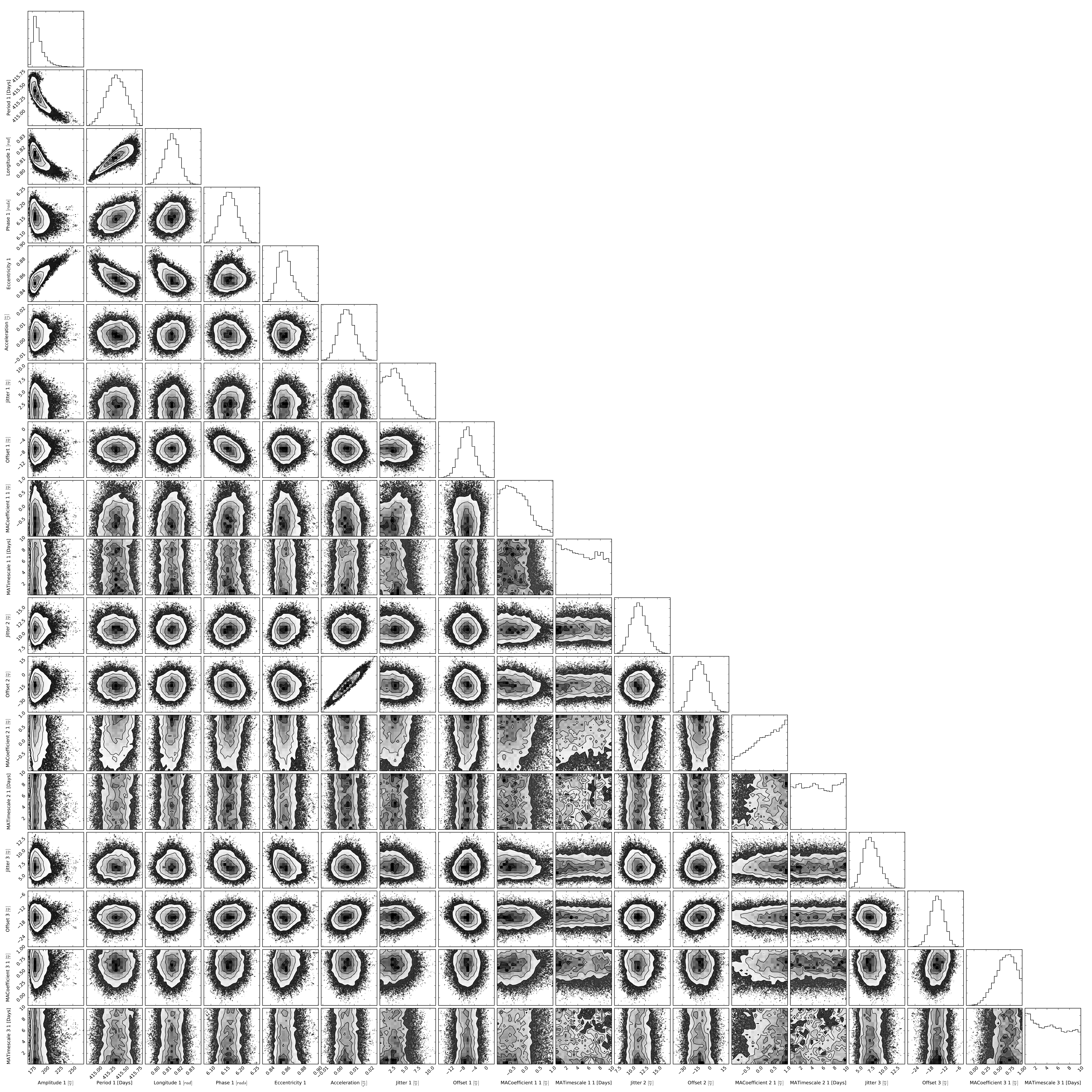}
\caption{Corner plot of the posterior distributions from the EMPEROR 
results for HD\,76920 single-planet fit.  These results are consistent 
with those obtained with \textit{Systemic}.}
\label{corner}
\end{figure}

\begin{figure}
\includegraphics[width=1.0\textwidth]{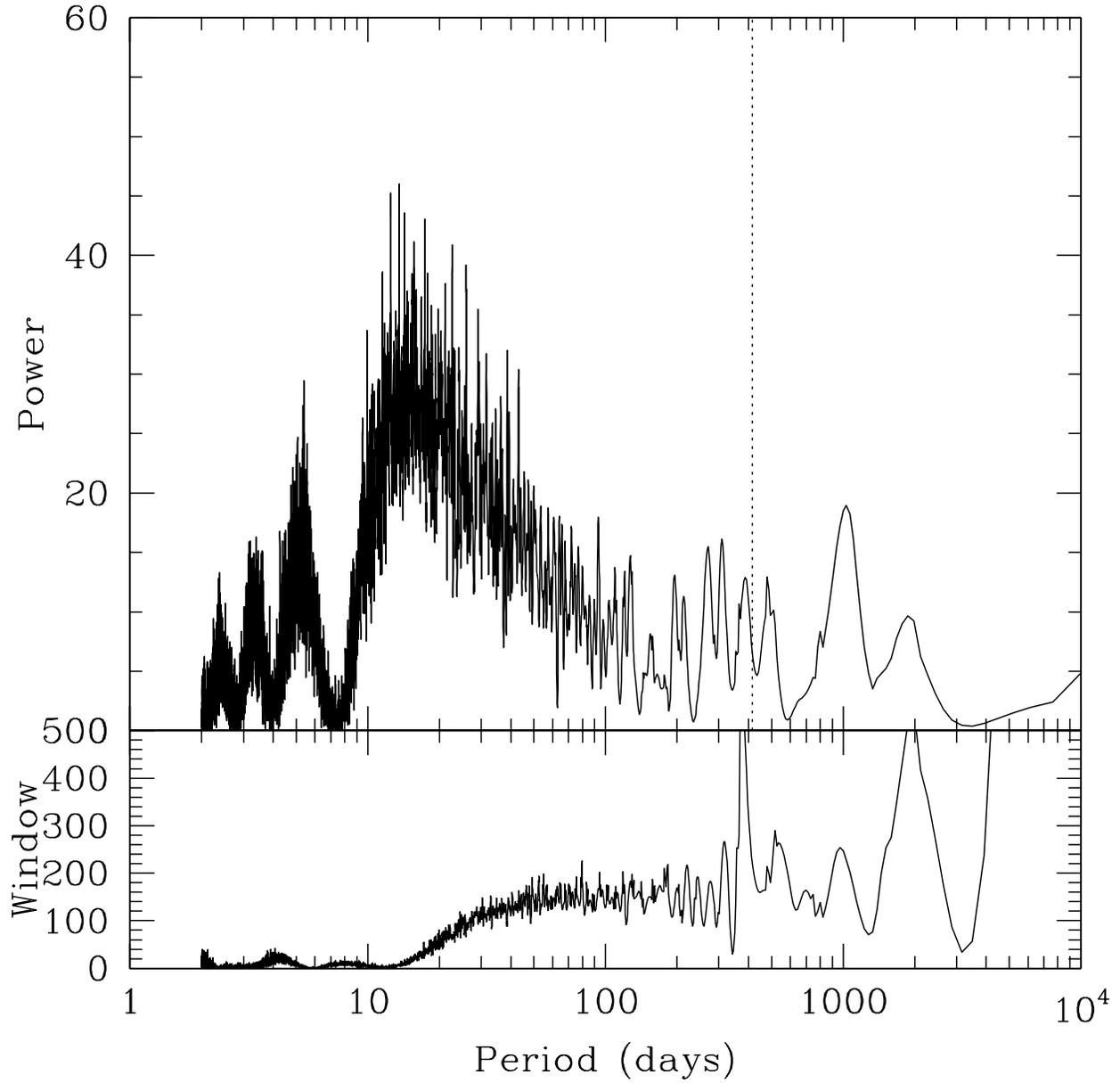}
\caption{Generalised Lomb-Scargle periodogram of ASAS photometry for 
HD\,76920.  A total of 1403 epochs spanning 8.8 years yield no 
significant periodicities.  The 415-day period of the planet is marked 
with a dotted line. }
\label{photpgram}
\end{figure}

\begin{figure}
\plottwo{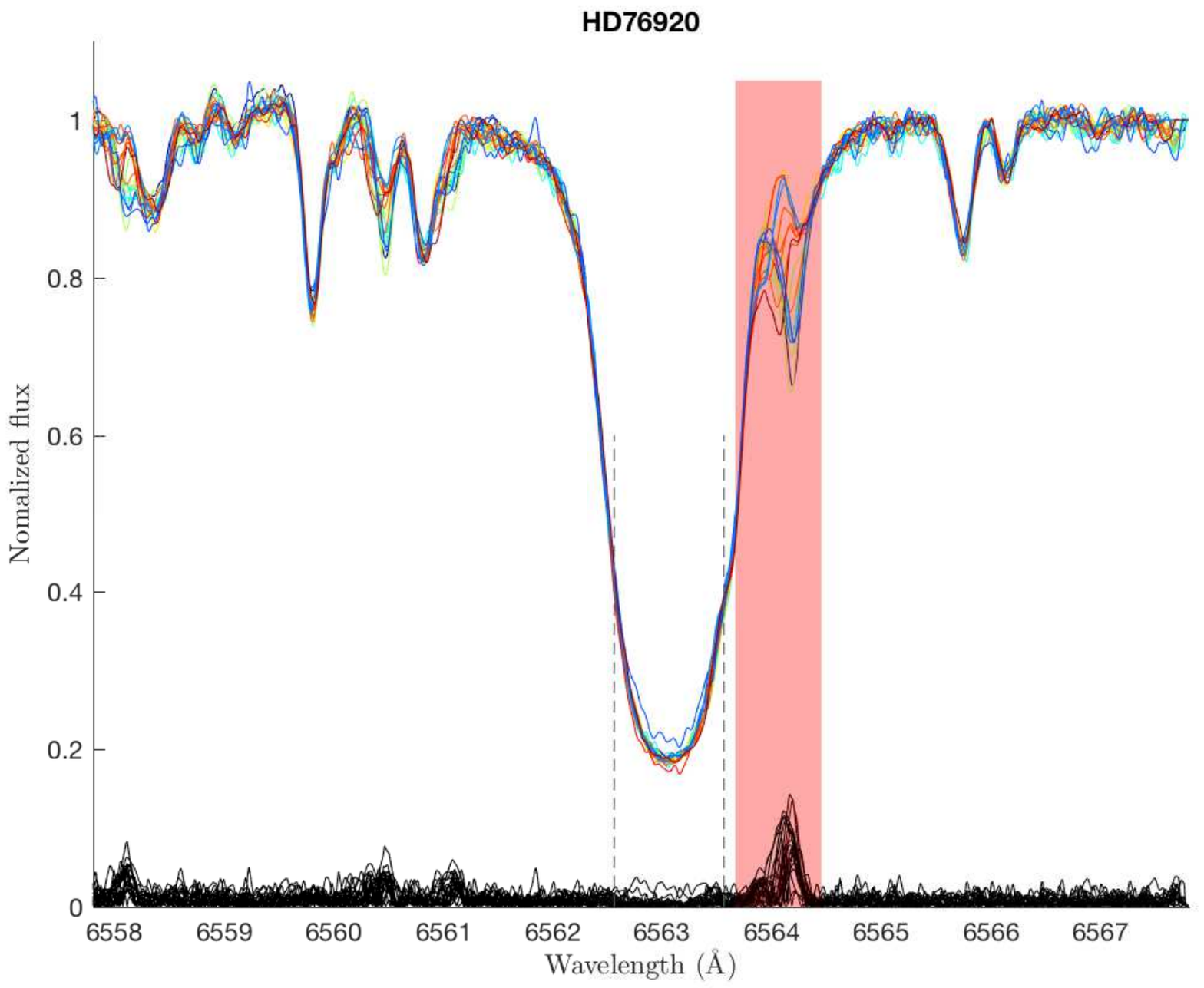}{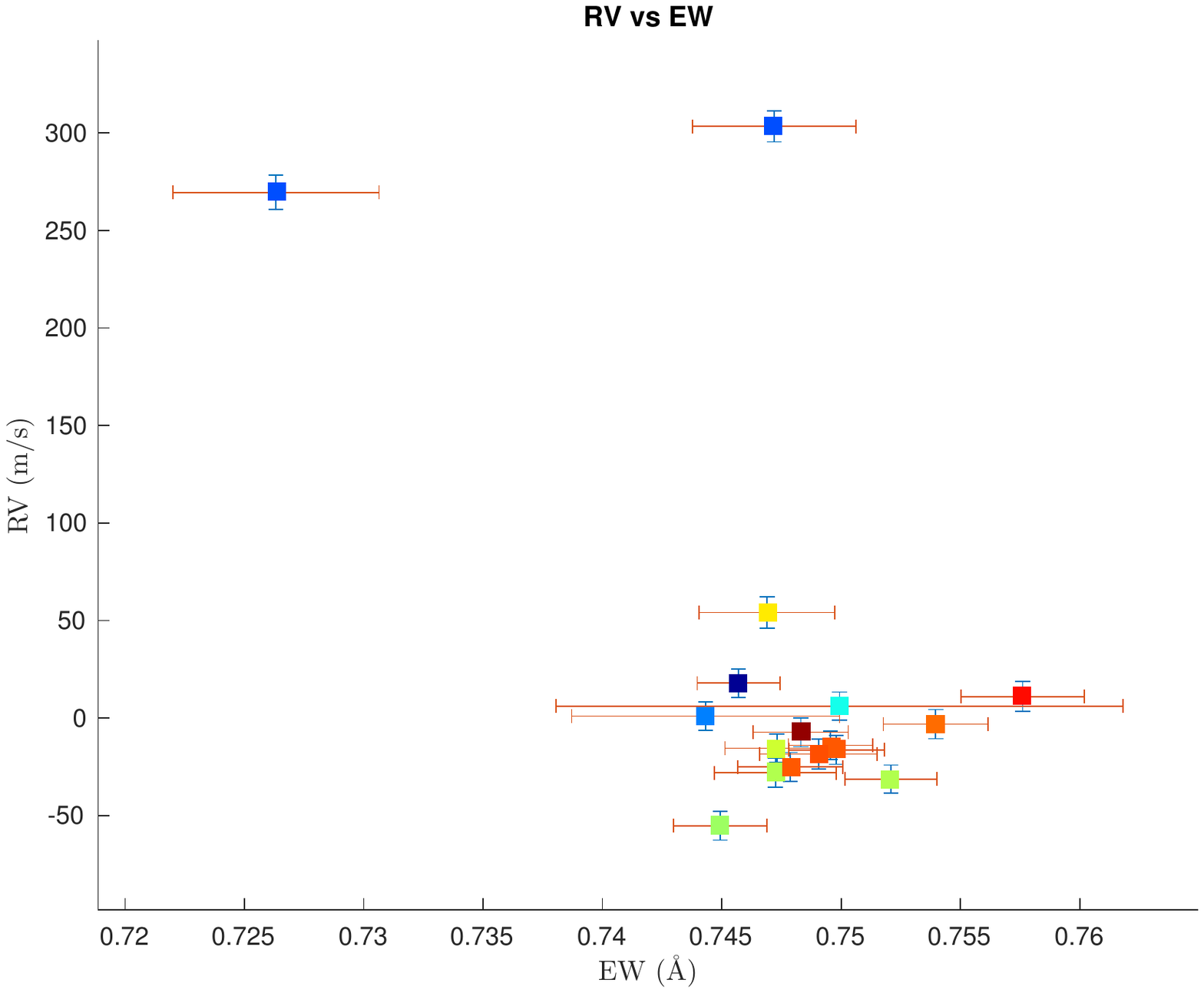}
\caption{\textit{Left:} the stacked normalised H$\alpha$ line from UCLES 
observations. The H$\alpha$ region is labelled within the black dashed 
lines, whereas the telluric region is highlighted in the red shaded 
area. Below are the residual amplitudes from the template (constructed 
as the weighted average of all observations). Large residuals are due to 
telluric contamination. \textit{Right:} radial velocity versus H$\alpha$ 
equivalent width. The same epochs are presented in identical colours 
across these two panels, and the closeness in colours within the same 
panel represents the closeness in BJD.}
\label{activity}
\end{figure}

\begin{figure}
\plotone{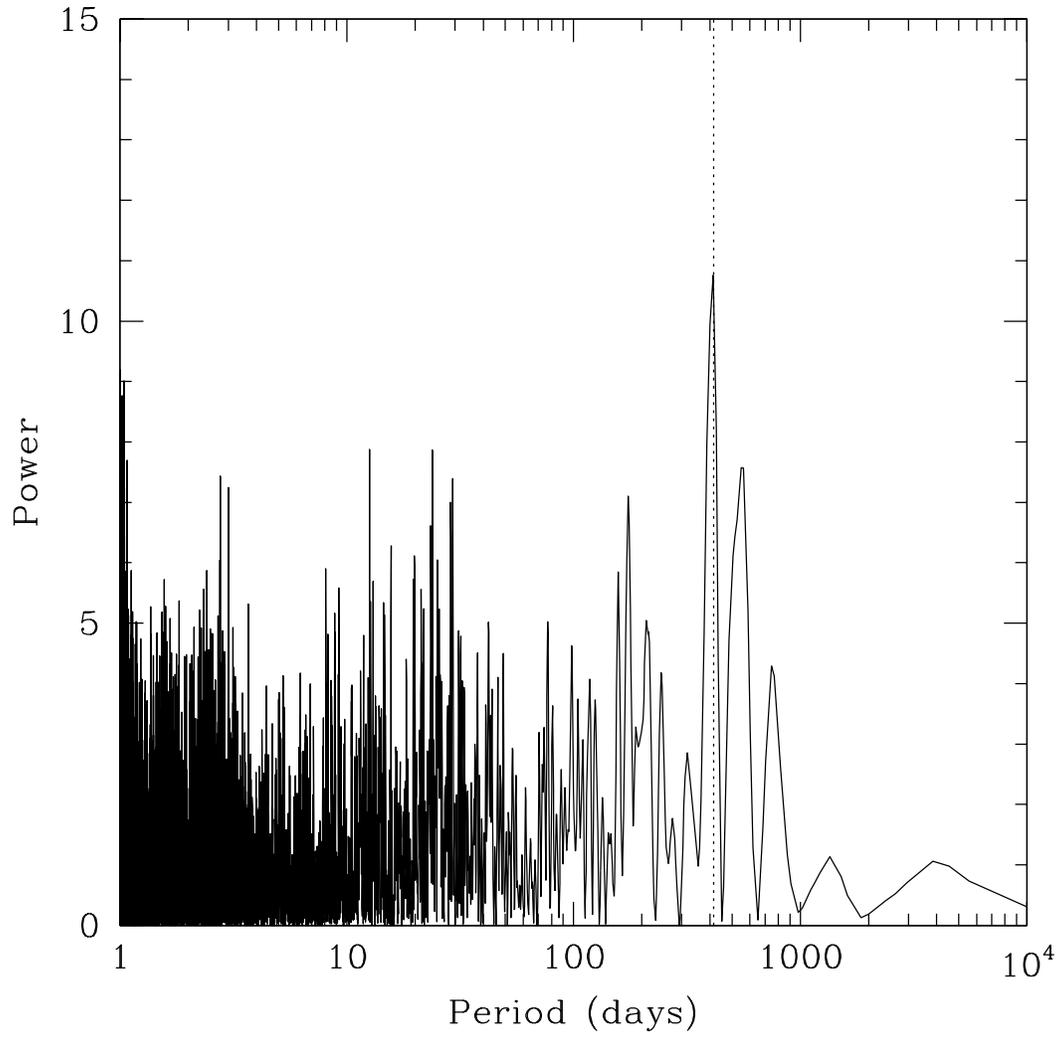}
\caption{Periodogram of the radial velocity data for HD\,76920 with the 
three velocity extrema removed.  The signal of the planet remains, 
with a false-alarm probability of 0.6\%.} 
\label{76920bAbides}
\end{figure}

Figure~\ref{howclose} shows the periastron distance for 116 confirmed 
planets\footnote{\url{http://exoplanets.org}, accessed 2017 May} 
orbiting giant stars, as a function of each planet's host-star radius.  
We note that the distinct absence of planets in the upper right quadrant 
of Figure~\ref{howclose} (i.e. highly eccentric planets that do not make 
particularly close approaches) is most likely an observational bias.  
That is, a highly eccentric planet on such an orbit would exhibit a 
radial velocity curve that is comparatively flat for most of the orbital 
cycle (e.g. Figure~\ref{fitplot}).  Such a target would be downgraded in 
observing priority, further diminishing the probability of catching the 
large velocity excursion that reveals the planet's existence.  

HD\,76920b moves on an orbit that brings the planet within $\sim$5 
stellar radii of its host star (i.e. 4 stellar radii from the surface).  
While this is a close approach, it is not the closest known; that honour 
falls to 4 UMa b \citep{dollinger07} which comes in to about 2 stellar 
radii of the surface of its host.  The estimated radii for these evolved 
stars are model-dependent and are fraught with uncertainties not 
reflected in Figure~\ref{howclose}.  Hence, the exact values are less 
important than the overall message, which is that highly eccentric 
planets orbiting evolved stars make close approaches and are thus 
valuable laboratories for studying star-planet interactions.

The origin of highly eccentric planets is often attributed to the 
Kozai-Lidov mechanism \citep{kozai, lidov}, whereby a binary stellar 
companion orbiting at $i\gtsimeq$39\degrees\ relative to the planet 
exchanges eccentricity and inclination with the planet, driving large 
excursions in planetary eccentricity.  This is likely the case for the 
two best-known extremely eccentric planets: HD\,20782b 
\citep{jones06,ecc3} and HD\,80606b \citep{ecc2,ecc4}, both of which are 
in systems containing a binary stellar companion.  However, we see no 
evidence for such a companion in the HD\,76920 system: there is no 
residual radial velocity trend, and no candidate stellar companions are 
visible within 5 arcminutes.

\citet{fh16} studied the influence of Kozai-Lidov oscillations to 
explain the lack of warm Jupiters around evolved stars.  They found that 
such oscillations efficiently remove warm Jupiters, showing that by the 
time the expanding star reaches $R>5$\,\Rsun, no planet has survived 
engulfment while an identical constant eccentricity population survives 
beyond 40\,\Rsun.  Although simulations of Kozai-Lidov oscillations are 
not available for the orbital distance of HD\,76920, the results of 
\citet{fh16}, and the fact that no stellar companion is found orbiting 
HD\,76920, suggest that Kozai migration is unlikely to be the origin of 
the observed eccentricity.  While capture of a free-floating planet is 
possible, such events typically emplace the captured body on very wide 
orbits.  For example, simulations by \citet{parker17} show that 
free-floating planets are exclusively captured onto orbits with 
$a>100$\,au.  It is unlikely that stellar perturbations could reduce the 
semi-major axis by 2-3 orders of magnitude.  A past episode of 
planet-planet scattering offers an alternative: high eccentricities can 
be attained in systems which eject one or more comparable-mass planets.  
In such systems, often a second planet is retained on a wide (10s-100s 
au) orbit \citep[e.g.][]{chatterjee08, mustill14, gotberg16}, which may 
escape detection in the current RV data.

\subsection{Transit probability}

If a highly eccentric gas giant happens to transit, it becomes all the 
more valuable, as it will offer a unique window into the physics and 
composition of ``cold Jupiters.'' At present only one such planet is 
known, HD\,80606b, with $e=0.93$ and an orbital period of 111 days.  It 
was discovered to transit in 2009 \citep{moutou09, fossey09, garcia09}, 
and the transit was further characterised in a multi-site ground-based 
observing campaign by \citet{winn09}.  A \textit{Spitzer} campaign 
centred on the periastron passage \citep{laughlin09} allowed the direct 
measurement of the atmospheric heating due to the $\sim$30-hour close 
approach.  For comparison, HD\,80606b passes to within 5.5 stellar radii 
of the star's surface at its closest approach (compared to 4 host-star 
radii for HD\,76920b).

The eccentricity of planetary orbits can have a major impact on the 
expected transit properties of the planet \citep{bar07,kan08}. The 
eccentric nature of the HD\,76920~b planetary orbit, combined with the 
relatively large size of the host star (see Table~\ref{stellarparams}), 
make this planet an intriguing prospect for transit observations. A 
similar case was studied by \citet{kan10} for the planet orbiting iota 
Draconis. In that case, the eccentricity is smaller, but the star is 
larger and the periastron passage of the planet occurs very close to 
inferior conjunction (where the true anomaly $f \sim 0\degr$). By 
contrast, the orbital fit from Table~\ref{planetparams} and the orbit 
visualization shown in Figure~\ref{orbit} demonstrate that inferior 
conjunction for HD\,76920~b occurs at a true anomaly of $f \sim 90\degr$. 
At this location in its orbit, the star--planet separation will be 
0.342~au, where the calculated orbital velocity of the planet will be a 
factor of 1.85 larger than the Earth's orbital velocity. The net effect 
of these factors is to produce a transit probability of 10.3\% and a 
transit duration of 2.3 days, assuming a Jovian planetary radius. By 
comparison, if the planet were in a circular orbit with the same 
semi-major axis of $a = 1.1491$~au, then the transit probability would 
be 3.1\% and the transit duration would be $\sim$4 days. The relatively 
large transit duration make this a difficult observation from the 
ground, but the most difficult aspect is the small predicted transit 
depth of 0.02\% resulting from the large stellar radius. The combination 
of transit probability and depth means that transiting giant planets 
around giant stars are likely plentiful but few have been detected. 
Currently, the largest stars ($R_\star = 6.3 \ R_\odot$) for which a 
planet has been detected are Kepler-91 \citep{lil14} and 
TYC\,3667-1280-1 \citep{n16}. Precision space-based photometry of giant 
stars will provide valuable information for the mass-radius relationship 
of giant planets around evolved stars.  These opportunities will by 
provided by TESS \citep{ricker14} and CHEOPS \citep{broeg13}.

\subsection{Circumstellar matter}

\begin{figure}
\plotone{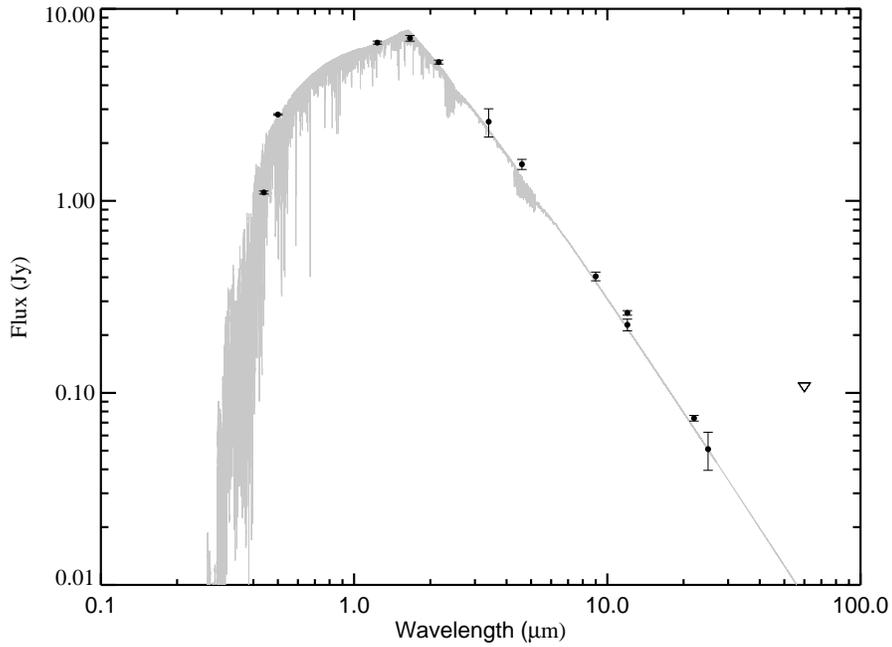}
\caption{Spectral energy distribution of HD\,76920.  The photometric 
data compiled from literature sources are shown as open black diamonds, 
with 1-$\sigma$ uncertainties.  The triangle at 60$\mu$m is an upper 
limit from IRAS.  The stellar photosphere model is shown in grey, and 
has been scaled according to the assumed stellar radius and 
parallax-derived distance (i.e. it is not a least-squares fit to the 
photometry).  No significant evidence of infrared excess is present, 
save for marginal 3$\sigma$ excesses in the WISE 3 and 4 bands. }
\label{sed}
\end{figure}

In \citet{ppps6}, we investigated the possibility of debris disks 
orbiting the giant stars HD\,86950 and HD\,29399, both of which were 
identified by \cite{mcdonald12} as having a possible infrared excess 
based on the presence of excess emission at $9\mu$m in the AKARI/IRC 
All-Sky Survey \citep{Ishihara2010}.  Since \citet{mcdonald12} also 
noted an infrared excess for HD\,76920 with a fractional luminosity 
($L_{dust}/L_{star}$) of $\sim 1.2\times\,10^{-3}$ peaking at $12\mu$m, 
we undertook a similar analysis in this work.  We compiled a spectral 
energy distribution from photometry spanning optical to mid-infrared 
wavelengths, including optical $BV$, near-infrared 2MASS $JHK_{s}$ 
\citep{Skrutskie2006}, WISE \citep{Wright2010}, AKARI 9~$\mu$m 
\citep{Ishihara2010}, and the IRAS faint source catalogue 
\citep{moshir90}.  We illustrate the stellar photospheric emission with 
a model from the BT-SETTL/Nextgen \citep{allard12} stellar atmospheres 
grid appropriate for the spectral type (K0\,III; $T_{\rm eff} = 4700~$K, 
log $g$ = 3.0, [Fe/H] = 0.0), and scaled to the stellar radius and TGAS 
distance \citep{ppps5, gaia16}.  We colour corrected the AKARI and WISE 
flux densities assuming blackbody emission from the star.  The resulting 
spectral energy distribution is shown in Figure~\ref{sed}.  No 
significant evidence of infrared excess is present.  The infrared excess 
noted in \citet{mcdonald12} is based on AKARI 9$\mu$, IRAS 12$\mu$, and 
IRAS 25$\mu$ data points.  We have added WISE 3 and WISE 4 photometry to 
that mix.  No evidence of significant excess from the target is present 
after colour correction of data points (IRAS12, IRAS25 have $K=1.4$ for 
a 5000\,K blackbody) and the calibration uncertainties of WISE 3 and 4 
($\sim$5\% and 6\%, respectively) are taken into account.

\subsection{Tidal effects and doomed worlds}

Planet--star tidal interactions become very strong when stars leave the 
main sequence. The increase in stellar radius means that the planet's 
gravity can more easily deform the star, and the star's deep convective 
envelope is highly efficient at dissipating the energy required for this 
deformation. The result is a damping of the planet's orbital semi-major 
axis and eccentricity. The dominance of the tide raised on the star, and 
the large stellar moment of inertia, mean that the planetary semi-major 
axis and eccentricity can continue to decay until the star engulfs the 
planet. This contrasts with the case of an eccentric planet orbiting a 
main-sequence star, for which the tide raised on the planet usually 
dominates, and the eccentricity decays to zero at a non-zero semi-major 
axis (see Fig.~8 of Villaver et al. 2014). Engulfment of the planet by 
the star is also aided by the rapidly expanding stellar radius. Working 
against this, as the star ascends the red giant branch (RGB), stellar 
mass loss begins to accelerate, causing the planet's orbit to expand. 
The fate of the planet thus depends on the stellar radius expansion, 
tidal forces dragging the planet inwards, and mass loss moving the 
planet out. The high eccentricity and modest semi-major axis of 
HD\,76920b mean that it is likely to be ingested by its host star as the 
latter ascends the RGB.

\begin{figure}
\plotone{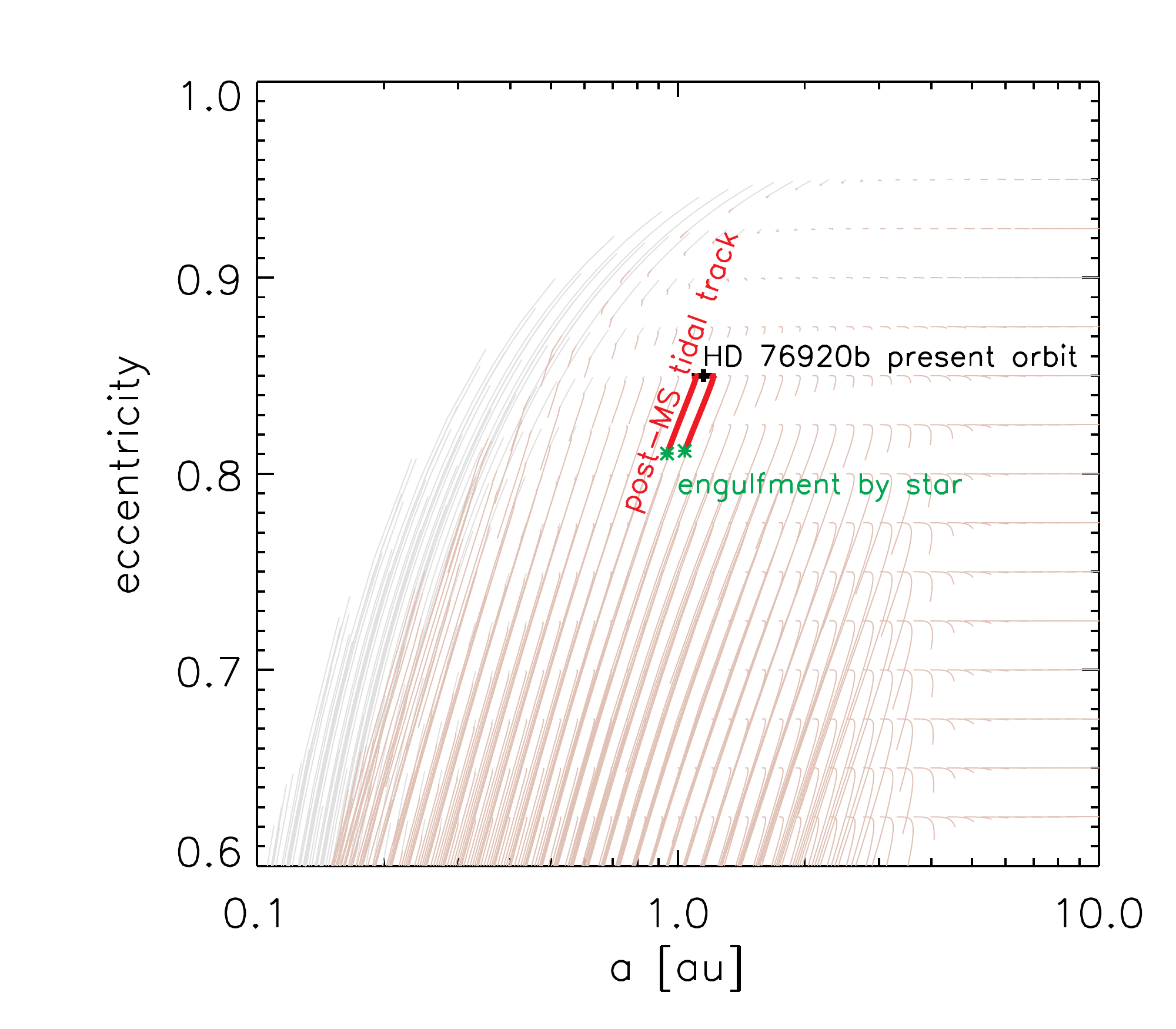}
\caption{Tidal evolution of planets orbiting a $1.17M_\odot$ star. 
Evolution along the main sequence is shown in grey, and evolution along 
the subgiant and RGB stages is shown in light brown. Two trajectories 
near the present location of HD\,76920b are highlighted. These end in 
engulfment by the swelling RGB star in around 100Myr.}
\label{tides}
\end{figure}

We model the future evolution of HD\,76920b using the method presented 
in \citep{villaver14}. This uses the tidal model of \citet{zahn77} for 
the tide raised on the star, which is suitable for highly-convective RGB 
stars, and \citep{matsumura10} for the tide raised on the planet. First 
we run a reference grid of planets at a range of semi-major axes (0.1--10 
au) and eccentricities (0--0.95) orbiting a $1.17M_\odot$ star (from 
SSE, Hurley et al 2000). The mass of the planets is 4\,\Mjup.  Their 
trajectories (after the first 1Gyr, during which planets at the top left 
circularise quickly) in $a-e$ space are shown in Figure~\ref{tides} as 
faint lines. Evolution along the main sequence is shown in grey, while 
evolution along the subgiant branch and RGB is shown in light brown. 
Many more planets are affected by tides on the RGB than on the MS. Two 
tracks starting close to the present orbit of HD\,76920b are 
highlighted. These tracks predict a modest decay of semi-major axis and 
eccentricity before the planet is engulfed in a little under 100Myr.  
Figure~\ref{howclose} shows the periastron distance in terms of the 
stellar radius. The periastron of HD\,76920b is at 4.82 times the 
stellar radius. Note that at $r_p/R_* \approx 2-3$, planets are in 
jeopardy \citep{villaver14} since that is where the tidal force starts 
to dominate the orbital evolution. The star HD\,76920 still has to 
evolve a bit up the red giant branch in order to tidally catch the 
planet at periastron.

While this planet will certainly end up engulfed by its host star, being 
too close for stellar mass loss to win over tidal orbital decay, making 
an exact prediction of its future evolution is challenging, partly 
because of the uncertainties in modelling tidal forces, and partly 
because of uncertainties on the stellar and planetary mass. As an 
example, dropping the stellar mass to $1.1M_\odot$ results in much 
stronger eccentricity decay before engulfment, with eccentricities 
dropping to around 0.7 at the time of engulfment. The high sensitivity 
of stellar evolutionary timescales on stellar mass means that at lower 
stellar masses there is more time for tidal decay to work to shrink the 
orbit before planetary engulfment.

The tracks shown in Figure~\ref{tides} use solar metallicity, while 
HD\,76920 has [Fe/H]$=-0.11$.  This introduces another source of 
uncertainty in the calculation of the decay timescales. The evolution of 
the star at a slightly lower metallicity than the one computed here is 
equivalent to the evolution of a more massive stellar mass at solar 
metallicity. HD\,76920 would in that case evolve a bit faster than 
assumed in Figure~\ref{tides} and thus will move more quickly into the 
stellar envelope. Note as well that mass-loss is expected to be affected 
by the metallity of the star and although red giant mass-loss rates are 
not very prominent they still have an effect in the final outcome of 
planetary systems (see Villaver et al. 2014).

\subsection{Future work}

Fortuitous observations have enabled us to characterise the orbit 
HD\,76920b as being unambiguously eccentric.  That is, the values of $e$ 
and $\omega$ have produced a radial velocity curve that cannot be 
mimicked by two low-eccentricity planets, a pathology that is distinctly 
possible when observations are sparse and poorly-sampled 
\citep[e.g.][]{shen08, ang10, aaps22, songhu}.  It would of course be 
desireable to achieve better phase coverage of the critical velocity 
excursion at periastron passage, to obtain a more precise measurement of 
the radial velocity amplitude (and hence the planet's mass).  We predict 
the next such passage to occur on 2018 Jan 17 (BJD 2458136.3$\pm$0.5), 
with $\sim$30 days of significant acceleration on either side of the 
velocity maximum.  Interested observers with dedicated (e.g. MINERVA: 
Swift et al. 2015) or queue-scheduled (e.g. CHIRON) telescope resources 
are highly encouraged to make plans to characterise the orbit of 
HD\,76920b at that time.  \citet{endl06} used the Hobby-Eberly Telescope 
in this manner to make high-cadence measurements of the periastron 
passage of HD\,45350b, an $e=0.76$ planet exhibiting a radial velocity 
curve similar to that of HD\,76920b.  Likewise, high-cadence 
observations capturing the periastron passage of HD\,37605b enabled 
\citet{cochran04} to confirm that highly eccentric planet ($e=0.737$).  
These examples highlight the importance of flexibly-scheduled radial 
velocity observations for truly understanding the orbital properties of 
unusual planets such as HD\,76920b.


\acknowledgements

We gratefully acknowledge the efforts of PPPS guest observers Brad 
Carter, Hugh Jones, and Simon O'Toole.  AJM is supported by the Knut and 
Alice Wallenberg Foundation.  JSJ acknowledges support by Fondecyt grant 
1161218 and partial support by CATA-Basal (PB06, CONICYT).  EV 
acknowledges support from the Spanish Ministerio de Econom\'ia y 
Competitividad under grant AYA2014-55840P.  This work has made use of 
data from the European Space Agency (ESA) mission {\it Gaia} 
(\url{https://www.cosmos.esa.int/gaia}), processed by the {\it Gaia} 
Data Processing and Analysis Consortium (DPAC, 
\url{https://www.cosmos.esa.int/web/gaia/dpac/consortium}). Funding for 
the DPAC has been provided by national institutions, in particular the 
institutions participating in the {\it Gaia} Multilateral Agreement.  
This research has made use of NASA's Astrophysics Data System (ADS), and 
the SIMBAD database, operated at CDS, Strasbourg, France.  This research 
has made use of the Spanish Virtual Observatory 
(http://svo.cab.inta-csic.es) supported from the Spanish MINECO through 
grant AyA2014-55216.  This research has also made use of the Exoplanet 
Orbit Database and the Exoplanet Data Explorer at exoplanets.org 
\citep{wright11, han14}.

\software{MOOG (Sneden 1973), Kurucz ATLAS9 (Castelli \& Kurucz 2003), 
SSE (Hurley et al. 2000), emcee (Foreman-Mackey et al. 2013)}



\clearpage

\begin{deluxetable}{lrrr}
\tabletypesize{\scriptsize}
\tablecolumns{4}
\tablewidth{0pt}
\tablecaption{Radial velocities for HD\,76920}
\tablehead{
\colhead{BJD-2400000} & \colhead{Velocity (\ms)} & \colhead{Uncertainty
(\ms)} & \colhead{Instrument}}
\startdata
\label{76920vels}
54867.07428  &      17.9  &    2.2  & AAT \\
55226.21880  &     269.5  &    5.3   & AAT \\
55227.20104  &     303.4  &    3.7  & AAT \\
55318.89227  &       1.0  &    1.9  & AAT \\
55602.04422  &       6.1  &    1.9  & AAT \\
55880.22005  &     -55.3  &    2.3  & AAT \\
55906.11204  &     -31.3  &    1.8  & AAT \\
55907.19640  &     -28.1  &    2.6  & AAT \\
55969.07596  &     -15.5  &    2.1  & AAT \\
56088.86366  &      54.1  &    3.8  & AAT \\
56344.02991  &      -3.1  &    2.7  & AAT \\
56374.98803  &     -16.4  &    2.4  & AAT \\
56376.95955  &     -14.1  &    2.4  & AAT \\
56377.96197  &     -25.2  &    2.6  & AAT \\
56399.96882  &     -18.5  &    3.1  & AAT \\
56530.31941  &      11.0  &    3.0  & AAT \\
56744.98572  &      -7.3  &    2.4  & AAT \\
\hline
57306.82770  &     311.8  &    4.4  & CHIRON \\
57324.78910  &      36.3  &    4.5  & CHIRON \\
57365.78950  &     -14.1  &    4.1  & CHIRON \\
57433.69900  &     -44.8  &    3.6  & CHIRON \\
57433.71310  &     -47.5  &    3.4  & CHIRON \\
57433.72720  &     -41.0  &    3.4  & CHIRON \\
57458.68830  &     -34.6  &    3.3  & CHIRON \\
57458.70240  &     -43.8  &    3.7  & CHIRON \\
57458.71650  &     -43.6  &    3.6  & CHIRON \\
57478.64630  &     -27.8  &    4.1  & CHIRON \\
57478.66040  &     -28.5  &    3.7  & CHIRON \\
57478.67450  &     -22.3  &    3.7  & CHIRON \\
\hline
 57641.91300  &     -36.9  &    5.0  & FEROS \\
 57643.90570  &     -24.2  &    5.4  & FEROS \\
 57700.84340  &      23.1  &    5.9  & FEROS \\
 57702.86840  &      20.3  &    4.3  & FEROS \\
 57703.79730  &      32.8  &    4.9  & FEROS \\
 57705.85330  &      38.2  &    5.0  & FEROS \\
 57894.56040  &     -30.7  &    6.2  & FEROS \\
 57895.46980  &     -22.7  &    5.5  & FEROS \\
\enddata
\tablecomments{The velocities shown are relative to instrument-specific
zero points, which are free parameters in the fitting process and are
given in Table~\ref{planetparams}.}
\end{deluxetable}

\clearpage

\begin{deluxetable}{lll}
\tabletypesize{\scriptsize}
\tablecolumns{3}
\tablewidth{0pt}
\tablecaption{Stellar Parameters for HD 76920}
\tablehead{
\colhead{Parameter} & \colhead{Value} & \colhead{Reference}
 }
\startdata
\label{stellarparams}
Spec.~Type & K1 III & \citet{houk75} \\
Distance (pc) & 184.8$\pm$7.5 & \citet{gaia16} \\
$(B-V)$ & 1.11$\pm$0.02 & \citet{hog00} \\
$E(B-V)$ & 0.0248 & \\
$A_V$ & 0.0769 & \\
Mass (\Msun) & 1.17$\pm$0.20 & \citet{ppps5} \\
$[Fe/H]$ & -0.11$\pm$0.10 & \citet{ppps5} \\
$T_{eff}$ (K) & 4698$\pm$100 & \citet{ppps5} \\
             & 4748 & \citet{mcdonald12} \\
             & 4744 & \citet{bj11} \\
log $g$ & 2.94$\pm$0.15 & \citet{ppps5} \\
Radius (\Rsun) & 7.47$\pm$0.6 & \citet{ppps5} \\
Luminosity (\Lsun) & 24.0  & \citet{ppps5} \\
                  & 21.7 & \citet{mcdonald12} \\
Age (Gyr) & 7.10 & \citet{ppps5} \\
\enddata
\end{deluxetable}


\begin{deluxetable}{lrr}
\tabletypesize{\scriptsize}
\tablecolumns{2}
\tablewidth{0pt}
\tablecaption{Keplerian orbital solution for HD 76920b}
\tablehead{
\colhead{Parameter} & \colhead{Value} }
\startdata
\label{planetparams}
Period (days) & 415.4$\pm$0.2 \\
Eccentricity &  0.856$\pm$0.009 \\
$\omega$ (degrees) &  352.9$^{+1.9}_{-1.1}$ \\
Mean anomaly\tablenotemark{a} (degrees)  & 46.5$\pm$0.4 \\
$K$ (\ms) &  186.8$\pm$7.0 \\ 
m sin $i$ (\Mjup) &  3.93$^{+0.14}_{-0.15}$ \\
$a$ (AU) &  1.149$\pm$0.017 \\
RMS about fit (\ms) &  9.74 \\
\hline
Zero point -- AAT \ms\ & 7.0$\pm$3.6 \\
Zero point -- CHIRON \ms\ & -23.7$\pm$4.9 \\
Zero point -- FEROS \ms\ & -13.5$\pm$4.7 \\
\enddata
\tablenotetext{a}{At epoch BJD 2454867.07428}
\end{deluxetable}

\clearpage

\begin{deluxetable}{lrr}
\tabletypesize{\scriptsize}
\tablecolumns{3}
\tablewidth{0pt}
\tablecaption{Habitable Zone Boundaries for Planet Candidate Host Stars}
\tablehead{
\colhead{Signals} & \colhead{BIC} & \colhead{$\Delta$BIC (k,k-1)} }
\startdata
\label{bics}
k=0 & 587.61 & \nodata \\
k=1 & 319.85 & 267.76 \\
k=2 & 299.21 & 20.64 \\
\enddata
\end{deluxetable}


\end{document}